\documentstyle[prc,psfig,aps,twocolumn]{revtex}
\begin{document}
\twocolumn[
\hsize\textwidth\columnwidth\hsize\csname @twocolumnfalse\endcsname

\draft
\begin{center}
PHYSICAL REVIEW C 2002 (IN PRESS)
\end{center}
\title{Non-perturbative effects in a rapidly expanding quark-gluon plasma}
\author{A. K. Mohanty, and P. Shukla}
\address{ Nuclear Physics Division,
Bhabha Atomic Research Centre,\\
Trombay, Mumbai 400 085, India}

\author{Marcelo Gleiser}
\address{Department of Physics and Astronomy,
Dartmouth College,\\
Hanover, NH 03755, USA}

\maketitle

\begin{abstract}
Within first-order phase transitions, we investigate the pre-transitional
effects due to the nonperturbative, large-amplitude thermal fluctuations which can
promote phase mixing before the critical temperature is reached from above.
In contrast with the
cosmological quark-hadron transition,
we find that the rapid cooling typical of the RHIC and LHC
experiments and the fact that the quark-gluon plasma
is chemically unsaturated
suppress the role of non-perturbative effects at current collider energies.
Significant supercooling is possible in a (nearly) homogeneous state of
quark gluon plasma.

\end{abstract}
]

\narrowtext

\section {Introduction}

It is possible to model the gross general features of a
phase transition from a quark-gluon plasma (QGP)
to a
hadronic phase through a phenomenological potential with a scalar order
parameter \cite{IGNATIUS}.
Assuming the transition to be discontinuous, or first-order, as
suggested by some recent lattice QCD simulations \cite{LQCD},
the QGP is cooled to a
temperature $T_1$, where a second minimum appears, indicating the presence of
a hadronic phase. With further cooling, the two phases become degenerate at
the critical temperature $T_c$, with a free energy barrier
which depends on physical parameters
characterizing the system, such as the surface tension ($\sigma$) and the
correlation length ($\xi$). This general behavior models both the cosmological
quark-hadron phase transition and the production of a QGP during heavy-ion
collision experiments,
as those under way at the RHIC and planned for the LHC.
In the latter case, the plasma generated by the collision
expands and cools, relaxing back to the hadronic phase.
Recent interest has been sparked by the possibility that this
relaxation process is characterized by the formation of disoriented
chiral condensates (DCC), which are coherent pion condensates similar to the domains
typical of quenched ferromagnetic phase transitions \cite{RAJA,BOYAN}.
The nonequilibrium properties of this relaxation process and DCC formation
has also been studied as a first order chiral phase transition where
the supercooled
phase may naturally lead to a "quenched" initial condition \cite{SCAV}.

Recent work on the dynamics of weak first-order phase transitions have shown
that, in certain cases, it is possible to have nonperturbative,
large-amplitude fluctuations before the critical temperature is reached, which
promote phase mixing \cite{GLE1}. Studies performed
in the context of the cosmological electroweak phase transition \cite{GLE2}
and quark-hadron phase transition \cite{SHUK} have indicated that, for a range
of physical parameters controlling the transition, these effects are present. It
is thus natural to consider if similar effects are present during heavy-ion
collisions \cite{AGAR}.

Whenever pre-transitional phenomena are relevant, one should expect
modifications from the usual homogeneous nucleation scenario, which is based
on the assumption that critical bubbles of the hadronic phase appear within a
homogeneous background of the QGP phase. The dynamics of weakly first-order
transitions will be sensitive to the amount of phase mixing at $T_c$: for
large phase mixing, above the so-called percolation threshold,
the transition may proceed through percolation of the
hadronic phase, while for small amounts of phase mixing, by the
nucleation of critical bubbles in the (inhomogeneous) background of isolated
hadronic domains, which
grow as $T$ drops below $T_c$.
An ideal quark gluon plasma in one dimension expands according to the Bjorken
scaling,
where $T^3t$ is constant \cite{BJOR}. Assuming the initial temperature
of the plasma produced at RHIC and LHC energies to be 2 to 3 times $T_c$,
scaling implies that the time ($\Delta t)$ taken by the plasma to cool from
$T_1$ to $T_c$  is of order  a few fm/$c$, which could be comparable with the
time scale of the subcritical hadronic fluctuations.
On the other hand, the expansion
rate of the early universe in the range $T_1 \le T \le T_c$ is
slow enough \cite{FULL,UNI}
($\Delta t $ could be of the order of a few $\mu$ secs), that
nonperturbative thermal
fluctuations may achieve equilibrium.
Another difference
is that collisions at RHIC and LHC
energies will lead to the formation of a highly (chemically) unsaturated
plasma, {\it i.e.}, the initial gluon and quark contents of the plasma remain
much below their equilibrium values \cite{SHUR,WANG,ESKOLA}.
A  chemically unsaturated plasma
will cool at an even faster rate than what is predicted from Bjorken
scaling \cite{BIRO,DUTTA}.
The cooling rate will also be accelerated further if expansion in
three dimensions is considered. Therefore, we will show that although
the equilibrium density distribution of
subcritical
hadron bubbles is significant - particularly when the transition is weak -
unlike the situation in cosmology, they do not contribute strongly
to phase mixing. For the range of parameters we investigated, of relevance for
RHIC and LHC energies, the plasma cools so rapidly that the subcritical
bubbles do not have time to reach their equilibrium distribution and promote
substantial phase mixing: significant supercooling is possible in a (nearly)
homogeneous quark gluon state.

\section {Sub-critical bubble formalism}

To study the dynamics of a first order phase transition,  we use a
generic form of the potential in terms of a real scalar order parameter $\phi$
given by \cite{IGNATIUS,SHUK},
\begin {eqnarray} \label {vphi}
V(\phi)=a(T)\phi^2-b~T \phi^3+c\phi^4.
\end{eqnarray}
The parameters
$a,~ b$ and $c$ are determined
from physical quantities, such as the
surface tension $(\sigma)$ and the correlation length ($\xi)$ of the
fluctuations,
and also from the requirement that the second
minimum of the above potential should be equal to the
pressure difference between
the two phases \cite{SHUK}. The bag equation of state is used to
calculate the pressure in the two phases. The potential $V(\phi)$
has a minimum at $\phi=0$ and a metastable second
minimum at
\begin{equation}
\phi_+=\frac{3bT+\sqrt{9b^2T^2-32ac}}{8c}
\end{equation}
below $T \le T_1$.
In the thin wall approximation \cite{LIN}, $b,~ c$ and $T_1$ can be
written as \cite{SHUK},
\begin {eqnarray} \label {con}
b=\frac{1}{\sqrt{6 \sigma \xi^5 T_c^2}};~c=\frac{1}{12 \xi^3 \sigma};
~T_1=\left[\frac{BT_c^4}{B-\frac{27}{16}V_b}\right]^{\frac{1}{4}},
\end{eqnarray}
where $B$ is the bag constant and
\begin{equation}
V_b(\phi_m)=\frac{3\sigma}{16\xi (T_c)}
\end{equation}
is the height of the degenerate
barrier at $T=T_c$ or at $a(T_c)=b^2T_c^2/4c$.
A wide spectrum of first-order phase
transitions, ranging
from very weak to strong, can be studied by either changing
$\sigma$
or $\xi$ or both.
For example, for a fixed value of $\xi$, the strength of the transition is
controlled by $\sigma$, becoming very weak first
order or second order when $\sigma \rightarrow 0$.

We follow Ref. \cite{GLE1} to obtain the equilibrium number density of
subcritical bubbles.
Let $n(R,t)$ be the number density of bubbles with
a radius between $R$ and $R+dR$
at time $t$ that satisfies the Boltzmann equation
\begin {eqnarray} \label {bolt}
\frac {\partial n}{\partial t} = -|v|\frac {\partial n}{\partial R} +
(1-\gamma)\Gamma_0 -\gamma \Gamma_+ .
\end{eqnarray}
The first term on the right-hand side is the shrinking term with velocity
$v=\partial R/\partial t$. The term $\Gamma_0$ is the rate per unit volume
for the thermal
nucleation of a bubble of radius $R$
of phase $\phi=\phi_+$ (hadron phase) within
the phase $\phi=0$ (QGP phase). Similarly, $\Gamma_+$ is the corresponding rate
of the phase $\phi=0$ within the phase $\phi=\phi_+$.  The factor $\gamma$
is defined as the volume fraction in the hadron phase.
Assuming $\Gamma_0 \approx \Gamma_+(=\Gamma)$
for a degenerate potential at $T=T_c$,
we write for the rate
\begin {eqnarray} \label {Gamma}
\Gamma = A T^4 \exp\left[-{F(\phi_+) \over T}\right].
\end{eqnarray}
where $A$ is a constant of order unity.
Using the Gaussian ansatz for subcritical
configurations;
\begin {eqnarray} \label {phi}
\phi(r)= \phi_+\exp\left(-{r^2 \over R^2}\right),
\end{eqnarray}
the free energy functional
\begin {eqnarray} \label {Fphi}
 F(\phi)=4\pi \int r^2 dr
\left [{1\over 2}({\partial \phi\over \partial r})^2 +V(\phi,T) \right]
\end{eqnarray}
can be written as \cite{GLE1}
\begin{equation}
F(\phi_+) =\alpha~R+\beta~R^3,
\end{equation}
where
\begin{equation}\alpha={3\sqrt{2}\pi^{3/2} \phi_+^2\over 8}
\end{equation}
and
\begin {equation}
\beta=\pi^{3/2}\phi_+^2\left[{\sqrt{2}a\over 4}-
     {\sqrt{3}bT\phi_+ \over 9}+{c\phi_+^2 \over 8}\right]
\end{equation}
The equilibrium number density $(n_0)$ of subcritical bubbles is found by solving
Eq.~(\ref{bolt}) with
$\partial n/\partial t=0$ and imposing the physical boundary condition
$n(R\rightarrow \infty)=0$. Using $\gamma_0
\approx 4\pi R^3n_0/3$, we get a coupled
equation for $\gamma_0$, which can be solved to get
\begin{eqnarray}\label{gamma_0}
\gamma_0=\frac{I}{1+2I}.
\end{eqnarray}
where
\begin{equation}
I=\int_R^\infty
\frac{4\pi}{3v}R^3 \Gamma(R^{'},\phi_+) dR^{'}~.
\end{equation}
We will consider the statistically dominant
fluctuations with $R \approx \xi$
and estimate $\gamma_0$
integrating Eq.~(\ref{gamma_0}) from $\xi$ to $\infty$.
Neglecting the shrinking term in Eq.~({\ref{bolt}),
the time dependent solution of $n(\xi,t)$
can be written as \cite{GLE1}
\begin {eqnarray} \label {nt}
n(\xi,t)=n_0(\xi)[1-\exp\{-q(\xi)t\}],
\end{eqnarray}
where
$q(\xi)=[(8\pi \xi^3/3)\Gamma]$ and
$n_0(\xi)=\Gamma(\xi)/q(\xi)$.
Alternatively, in term
of $\gamma$, the above solution has the form
\begin{eqnarray}\label{gamma}
\gamma(\xi,t)=\gamma_0(\xi)[1-\exp(-q_0t)],
\end{eqnarray}
where $q_0=(4\pi \xi^3/3)\Gamma/\gamma_0$. The relaxation time
$\tau=q_0^{-1}$ depends on two factors $\gamma_0$ and $\Gamma$ out of which
only the $\gamma_0$ is affected by shrinking (if included). Since
we know the complete solution of $\gamma_0$ that includes shrinking
[Eq.~(\ref{gamma_0})], Eq.~(\ref{gamma}) can also be used to estimate its
time dependence. Note that
the presence of a shrinking term in Eq.~(\ref{bolt}) results in a reduction of
$\gamma_0$ and also in a faster relaxation process.

\section {Result and Discussion}

 First we consider the slow evolution of the medium as in the case of early
universe \cite{FULL,UNI} so that the equilibrium scenario is applicable.
Figure~1 shows the plot of $\gamma_0$ as a function of $\sigma$ at $T=T_c$
for a few typical values of the prefactor $A$.
We have fixed $\xi$ at 0.5 fm, $T_c$ at 160 MeV and $v=1/\sqrt{3}$.
As expected,
the equilibrium hadronic fraction increases
with decreasing $\sigma$ and becomes as large
as 0.5 for weak transitions.
Recent lattice QCD predictions \cite{LQCD}
suggest that the quark-hadron phase transition
could be weakly first order with  $\sigma$ values in the range
2 - 10 MeV/fm$^2$.
Therefore, the choice of
$\sigma$ in the above range and
$A\sim 1$ \cite{LIN} would imply significant amount of phase
mixing at $T=T_c$ so that homogeneous
nucleation becomes inapplicable \cite{SHUK}.

\begin{figure}
\centerline{\psfig{figure=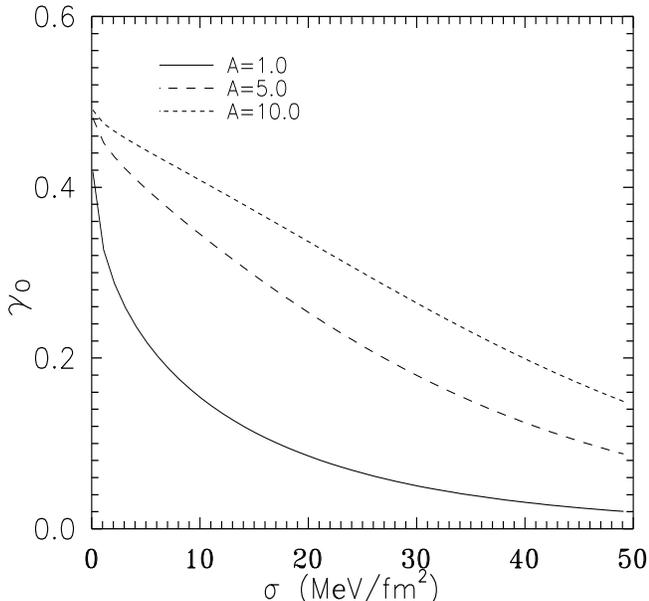,width=8.5cm,height=8cm}}
\caption{$\gamma_0$ versus $\sigma$ at $T=T_C$ for a few
typical values of $A$. $\xi$ is fixed at 0.5 fm and $T_C$ at 160 MeV.}
\label{fm1}
\end{figure}

Next we consider the plasma expected to be formed at RHIC and LHC energies.
Since the expansion of such plasma is much faster compared to the plasma
at the early universe,
it will be interesting to know the amount of phase mixing (the value of
$\gamma$) built up
by the time the plasma cools from $T_1$ to $T_c$.
Assuming ideal scaling,
we can estimate the time $\Delta t$ taken by the plasma to cool from $T_1$
to $T_c$ as
\begin{eqnarray} \label{tim} \Delta t =
\frac{T_0^\nu}{T_c^\nu}t_0 \left[1-\frac{T_c^\nu}{T_1^\nu}\right]~,
\end{eqnarray}
where $\nu=3$ in (1+1) dimensions.
Since $T_1$ depends on $\sigma$ [see Eq.~(\ref{con})],
$\Delta t$ will also depend on $\sigma$, being smaller the
weaker the transition.
In the standard scenario, we can assume the initial temperature
$T_0 \approx$ 320 MeV and the formation time $t_0 \approx1$ fm.
However, several perturbative-inspired QCD models \cite{WANG,ESKOLA,GEI} 
suggest
a very different collision scenario at RHIC and LHC energies, which lead to the
formation of unsaturated plasma with high gluon content. Such a plasma will
attain thermal equilibrium in a short time $t_0 \approx 0.3 - 0.7$ fm, but
will remain far from chemical equilibrium. Since the initial plasma is
gluon rich, more quark and anti-quark pairs will be needed in order to
achieve chemical equilibration. The dynamical evolution of the plasma undergoing
chemical equilibration was studied initially by Biro et. al. \cite{BIRO} and
subsequently by many others \cite{DUTTA} by solving  the hydrodynamical
equations along with a set of rate equations governing chemical equilibration.
It was found that a chemically unsaturated plasma cools faster than what is
predicted by Bjorken scaling, since additional energy is consumed
during chemical equilibrium. Following Ref. \cite{DUTTA}, we have studied
chemical equilibration and dynamical evolution of the QGP with
two sets of initial conditions, HIJING \cite{WANG} and Self Screened Parton
Cascade Model (SSPM) \cite{ESKOLA},
as listed in table I.
The Perturbative QCD inspired models like Parton Cascade Model (PCM) \cite{GEI}
and
HIJING (Heavy Ion Jet Interaction Generator) \cite {WANG} are generally used to simulate
the nuclear collisions at collider energies on the level of microscopic
parton dynamics. The PCM calculations describe the space time evolution of
quark and gluon distributions by Monte Carlo simulations of relativistic
transport equations.  The HIJING model also incorporates  the perturbative
QCD approach and multiple minijet productions, however, it does not incorporate
a direct space time description. Early PCM calculations were done by assuming
a $p_T$ cutoff to ensure the applicability of the perturbative expansion of the
QCD scattering process. In the recently formulated self
screened parton cascade model (SSPM) [3],
early hard scattering produces a medium which screens the longer range
color fields associated with softer interactions. The screening occurs
on a length scale where perturbative QCD still applies and the divergent
cross sections in the calculation of the parton production can be regulated
self-consistently without an ad hoc cutoff parameter. The numerical
studies based on the parton cascade model suggest that the parton plasma
produced in the central region is essentially a hot gluon plasma and the
dynamics is mostly dominated by gluons. Gluons thermalize rapidly
reaching approximately
isotropic momentum distributions in a very short time scale.
The densities of quarks and antiquarks stay well below the gluon density
and can not build up to the full equilibration values required for an ideal chemical
mixture of gluons and quarks. The similar conclusions
have also been drawn from the calculations based on HIJING approach. Though
both PCM and HIJING are QCD inspired models, the two still differ in
quantitative predictions possibly due to different treatment of multiple
parton interactions and collective effects. In the following, we take
the initial conditions obtained both from HIJING and SSPM calculations
at the time when parton momentum distribution
becomes isotropic.
We consider two dominant reaction
channels $q {\bar q} \rightleftharpoons gg$ and
$ gg\rightleftharpoons ggg$ that contribute to the chemical equilibrium.
The fugacity $\lambda_{g(q)} (\le 1)$ gives
the measure of the deviation of the gluon (quark) density from the equilibrium
value; chemical equilibrium
is achieved when $\lambda_i$'s $\rightarrow 1$. For a detail discussion on
chemical equilibration, we further refer to \cite{DUTTA}.

\begin{table}
\begin{center}
\caption{Initial conditions are taken from Ref.~[20]
as predicted by SSPM and HIJING calculations.
The fugacities $\lambda_i$'s give a measure of the deviation of the gluon or
quark densities from the equilibrium values.}
\begin{tabular}{|l|l|l|l|l|l|l|}
\hline
CODE&ENERGY & $t_i$ (fm/$c$) & $T_i$ (GeV) &$\lambda_g$ &$\lambda_q$ & $\nu$\\
\hline
SSPM&RHIC  &0.25        & 0.668       &0.34        &0.064      & 2.2 \\
\hline
SSPM&LHC   &0.25        & 1.02        &0.43        &0.082      & 2.2 \\
\hline
HIJING &RHIC   &0.7     & 0.55        &0.05        &0.008      & 1.9 \\
\hline
HIJING &LHC  &0.5     & 0.82            &0.124       &0.02       & 1.8 \\
\hline
\end{tabular}
\end{center}
\end{table}

Figure~2 shows
a typical example of the effect of chemical equilibration on the cooling
rate for SSPM initial conditions at RHIC energy ($\lambda_{g0}=0.34$,
$\lambda_{q0}=0.064$, $t_0=0.25$ fm and $T_0=0.668$ GeV). The dotted curve
(marked as $T_B$) shows the cooling rate as a function of time which obeys
Bjorken's scaling ($T^3t$=constant) corresponding to the case of an equilibrated
plasma ($\lambda_g=\lambda_q=1.0$). In case of a chemically unsaturated
plasma for which the values of initial fugacities are much less than unity,
the hydrodynamical expansion of the plasma proceeds along with chemical
equilibration. As a result, both $\lambda_g$ and $\lambda_q$ increase with time
as well as the temperature (shown by dashed curve) drops at a faster rate
as compared to the Bjorken's scaling. The solid circles show the
temperature given by ($T^\nu~t $=constant for $\nu=2.2$).
In this work, since we are interested
only in the cooling rate, we skip the details of the calculation
and parameterize the cooling rate in terms of
$\nu$ in the range $T_1 \le T \le T_c$ (i.e. $T^\nu t$=const).
In table I, $\nu$
has been listed for two sets of initial conditions obtained using HIJING
and SSPM models at RHIC and
LHC energies. Note that $\nu < 3$ implies a faster cooling.
Figure~3 shows the plot of $\Delta t$ as a function of $\sigma$
as obtained from Eq.~(\ref{tim}) for different $\nu$ values.
The time $\Delta t$ depends on the initial values of the temperature $T_0$,
formation time $t_0$ and also on the cooling rate $\nu$.
However, except for the SSPM initial conditions at LHC energies,values of
$\Delta t$ obtained with other initial conditions have nearly similar values.

\begin{figure}
\centerline{\psfig{figure=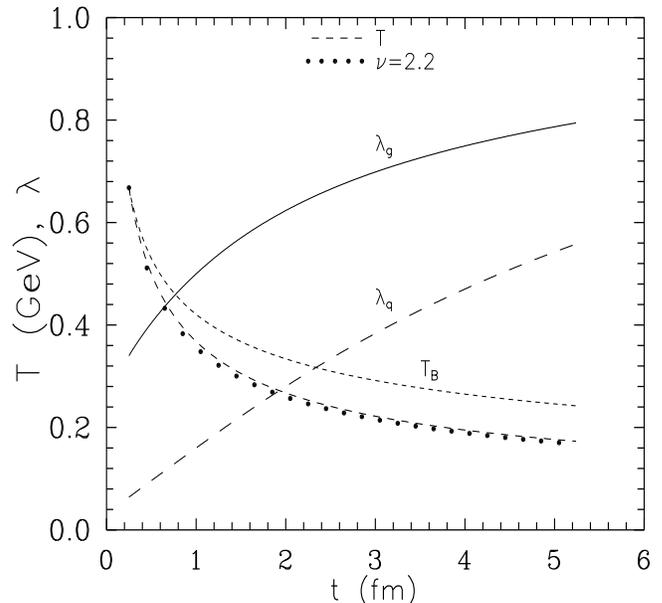,width=8.5cm,height=8cm}}
\caption{ The temperature $T$ and fugacity $\lambda$ as a function of time
    $t$. The description of the various curves are given in the text.}
\label{fm2}
\end{figure}

\begin{figure}
\centerline{\psfig{figure=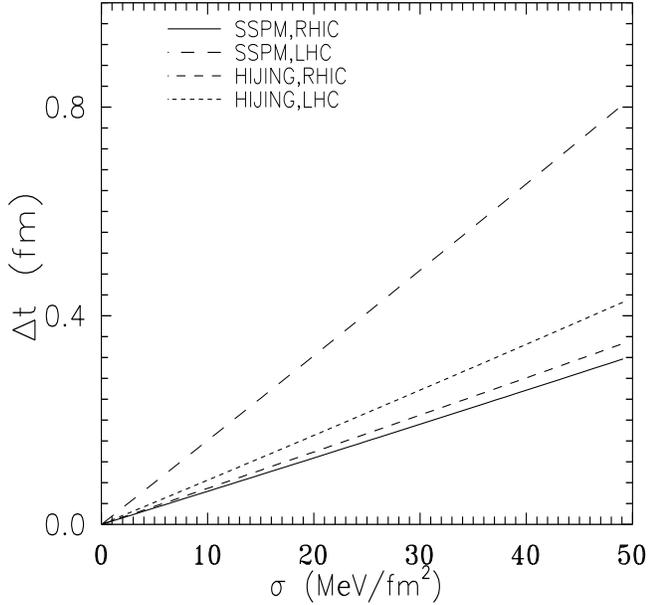,width=8.5cm,height=8cm}}
\caption{$\Delta t$ as a function of $\sigma$ for various initial
        conditions as shown in table I.}
\label{fm3}
\end{figure}

Next we proceed to estimate the density of subcritical hadron bubbles
built up at $t=\Delta t$.
Figure~4 shows $\gamma(t)/\gamma_0$ as a function of $t$ at three different
$\sigma$ values. The equilibration rate of the subcritical hadron bubbles
of a given radius depends on the ratio $\Gamma/\gamma_0$. Although both
$\Gamma$ and $\gamma_0$ are larger for weaker transitions,
their ratio decreases
with decreasing $\sigma$. Therefore,
as can be seen, equilibration is faster for a stronger transition
as compared to the weak one.

\begin{figure}
\centerline{\psfig{figure=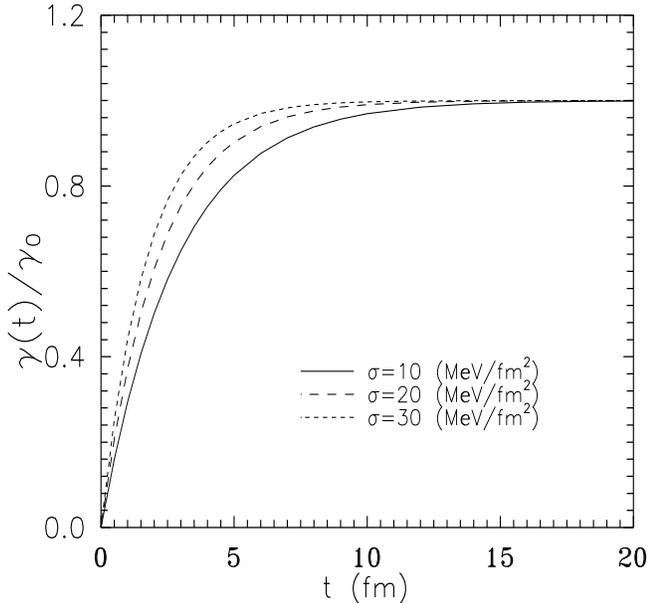,width=8.5cm,height=8cm}}
\caption{The ratio $\gamma(t)/\gamma_0$ as a function of $t$ at three
typical values of $\sigma$ for $A$=1.}
\label{fm4}
\end{figure}

Figure~5 shows the fraction of the density
built up at time $t=\Delta t$ as a function of $\sigma$ with different initial
conditions.
Although the equilibrium density distribution of subcritical hadron bubbles
increases with decreasing $\sigma$, the time $\Delta t$ decreases
with decreasing $\sigma$. As a result of these two competing effects,
$\gamma$ at $t=\Delta t$ shows
a peak at around $\sigma \approx$ 20 MeV/fm$^2$.
The equilibrium fraction $\gamma_0$ depends on the ratio $A/v$, which increases
either due to increase in $A$ or decrease in $v$. However, the variation in $A$
and $v$ act differently on $q_0$ as the nucleation rate $\Gamma$ depends only
on $A$. Therefore, we study the
effect of $A$ and $v$ on $\gamma_0$ and $\gamma$  separately.
  Figure~6(a)
shows the plot of $\gamma_0$ (upper curves) and $\gamma(t)$ (lower curves)
as a function of $\sigma$ at $A$=5,10 and 20 respectively. Other parameters
are $v=1/\sqrt{3}$, $T_c=160$ MeV and $\xi$=0.5 fm. As expected, $\gamma_0$
goes up as $A$ increases. The increase in $\gamma_0$ for $A$ from 5 to 20
is about
1.5 to 2 times, but the nucleation rate $\Gamma$ goes up by a factor of
4. Therefore, the ratio $\Gamma/\gamma_0$ also goes up resulting in a faster
equilibrium. The net consequence is both $\gamma_0$ and $\gamma(t)$ go up
with increasing $A$. For the calculation of $\gamma(t)$, we have used
SSPM and RHIC initial conditions. Further, we would like to mention
here that although we have varied $A$ up to 20, the value of $A$ more than
unity is unrealistic. A recent work by us \cite{AKM}  and also studies
in ref \cite{CSER} suggest $A<<1$. However, the ratio $A/v$ can also go up
with decrease in $v$, which
we study in figure~6(b).
Figure~6(b) shows $\gamma_0$ and $\gamma$ for $v=c=1$ (upper limit), $1/4$
and $1/12$. This corresponds to a $A/v$ ratio of 5, 20 and 60
respectively. Therefore, $\gamma_0$ goes up with decreasing $v$ as expected.
Since $A$ is fixed, $\Gamma$ does not change, but $q_0$ decreases with
increasing $\gamma_0$ resulting in slower equilibration. As a result,
$\gamma(t)$ does not build up at all. It is also interesting to note
that $\gamma(t)$ does not get affected much by the choice of $v$ although
$\gamma_0$ has a strong dependence on it. The $\gamma(t)$ only depends
on parameter $A$. This aspect is interesting.

\begin{figure}
\centerline{\psfig{figure=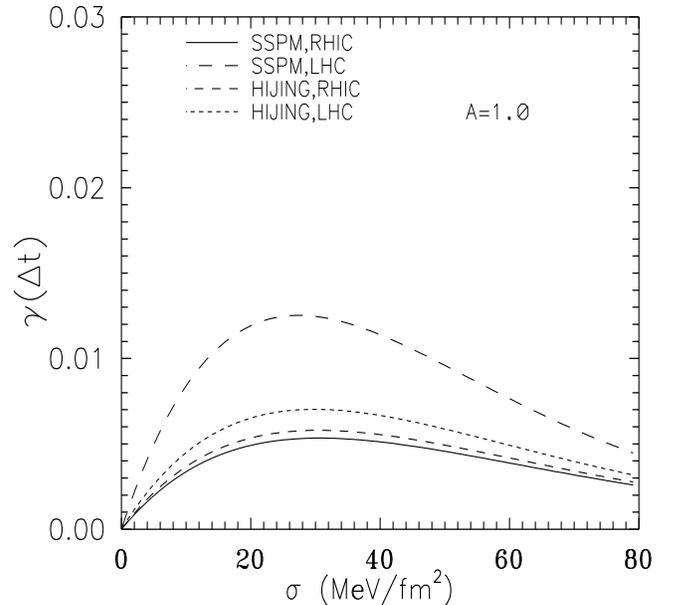,width=8.5cm,height=8cm}}
\caption{ The fraction $\gamma$ at $t=\Delta t$ as a
         function of $\sigma$ at $A=1$ and $v=0.577$.}
\label{fm5}
\end{figure}

\begin{figure}
\centerline{\psfig{figure=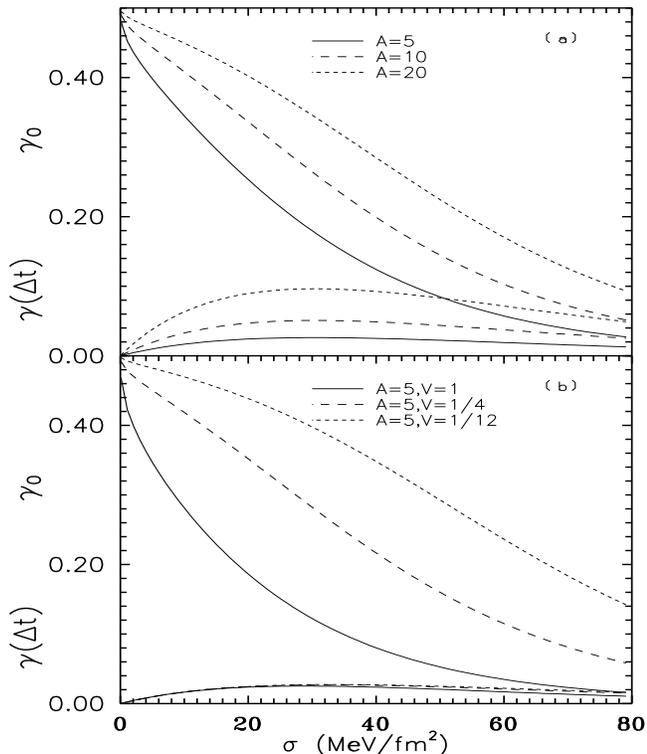,width=8.5cm,height=10cm}}
\caption{(a) The fraction $\gamma_0$ (upper curve) and $\gamma$ at
             $t=\Delta t$  (lower curve) as a
             function of $\sigma$ at $A=5, 10, 20 $ and $v=0.577$.
           (b) Same as above at $A$=5, but for different $v$ values with
           SSPM initial conditions at RHIC energy.}
\label{fm6}
\end{figure}

From the above studies (from figures~ 5 and 6),
we can conclude that,
the fraction in the range $2 {\rm MeV/fm}^2 \le \sigma \le 10 {\rm MeV/fm}^2$
does not build up to a significant level due to rapid cooling
of the plasma, although the equilibrium concentration is fairly large.
It may be mentioned here that we have considered expansion only in (1+1)
dimensions. Inclusion of transverse expansion, significant at RHIC and LHC
energies, will accelerate the cooling rate further,
reducing the amount of
phase mixing considerably.
Since
phase mixing at $T=T_c$ is negligible, the plasma will supercool
and the phase transition may proceed by the nucleation of critical-size
hadron bubbles within a (nearly) homogeneous background of the
metastable QGP phase.

We have also studied the effect of other parameters like $T_c$ and $\xi$
on $\gamma$.
Figure~7(a) shows the plots for various $T_c$ values at A=5.
The nucleation rate decreases with decreasing $T_C$ [see Eq.~(\ref{Gamma})]
resulting in a decrease in
$\gamma_0$. On the other hand, smaller $T_c$ will result in larger $\Delta t$,
which may increase $\gamma(t)$. However, as shown in figure~7(a), the variation
in $\gamma(t)$ with $T_c$ is not very significant although $\gamma_0$
depends on it. Similarly, figure~7(b) shows the plots at various $\xi$ (0.5, 1.0 and
2.0). Increasing $\xi$ suppresses $\gamma_0$ and $\gamma(t)$ particularly
when the transition is strong.
Therefore, the effect of other parameters like $v$, $T_c$ and $\xi$
on $\gamma$ are not very
significant to promote phase mixing. The prefactor $A$ is the only sensitive
parameter on which $\gamma(\Delta t)$ depends.
While the choice of $A\approx 1$ is quite reasonable \cite{LIN},
we have also varied $A$
from 1 to 20 and did not find significant phase mixing particularly
when $\sigma$ is small.

\begin{figure}
\centerline{\psfig{figure=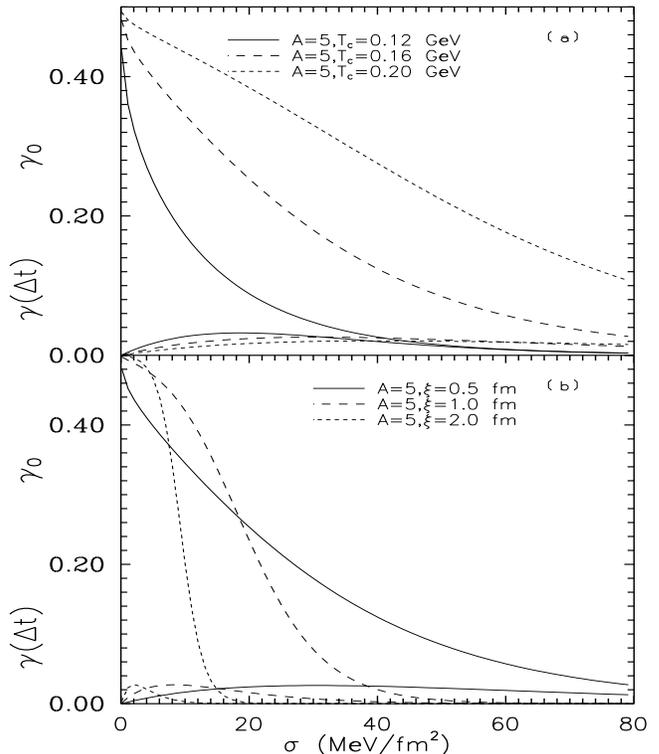,width=8.5cm,height=10cm}}
\caption{(a) The fraction $\gamma_0$ (upper curve) and $\gamma$ at
             $t=\Delta t$  (lower curve) as a
             function of $\sigma$ for different values of $T_C$
             at $A=5$ and $v=0.577$.
           (b) Same as above at $A$=5, but for different $\xi$ values with
           SSPM initial conditions at RHIC energy.}
\label{fm7}
\end{figure}

\section {Conclusion}

In conclusion, we have studied the effect of phase mixing promoted by thermal
subcritical hadron bubbles during a first-order quark-hadron phase
transition as predicted to occur during heavy-ion collisions.
Although the equilibrium density distribution of these subcritical
bubbles can be quite large, their equilibration time-scale is larger than
the cooling time-scale for the QGP. As a consequence, for RHIC and LHC
energies, they will not build up to a level capable of modifying the
predictions from homogeneous nucleation theory.
The phase transition may proceed either through the
nucleation
of critical size hadron bubbles in a (nearly) homogeneous background of
the supercooled quark-gluon plasma or through spinodal decomposition if nucleation
rate is not significant \cite{SCAV1}.
This situation is to be contrasted with the cosmological quark-hadron
transition, where substantial phase mixing may occur, altering the
dynamics of the phase transition.
We would also like to add here that even though
our calculations rule out the role of subcritical bubbles, it is possible
that impurities may increase the decay time-scale and no real supercooling
will be measured, as is the case with many condensed matter systems. The
question, however, remains as to what these impurities, if any, might be
in this context. One possibility -- ruled in this work -- is that the
subcritical bubbles, being seeds for nucleation, may act as 
impurities  \cite{GLE2}. However, other possibilities, as the
presence of condensates, may exist and should be considered in the near
future. If there is
supercooling there will be an extra entropy production which will reflect
on the final hadron multiplicities. In this case,
subcritical bubbles are not present, or are
irrelevant. On the other hand, if the transition is first order and no
extra entropy is observed, subcritical bubbles (or unknown impurities...) do
play a role.

\acknowledgements
We thank A. Dumitru for many fruitful comments and discussions.
MG acknowledges the partial support by National
Science Foundation Grant PHY-0070554.


\begin{thebibliography}{99}
\bibitem{UNI} The solution of Einstein's relativistic field equation
              yields a relation between the age of the universe and
              the temperature \cite{FULL},
$t=\sqrt{\frac{\alpha}{G}}T^{-2}$ where
              $\alpha=9/(164 \pi^3)$. Since the Newton's
              constant $G$ is very small, the above
              relation would imply

$\Delta t=\sqrt{\frac{\alpha}{G}}(T_c^{-2}-T_1^{-2}) \sim$ few $ \mu$ sec.
\end{thebibliography}
\end{document}